\pdfoutput=1 
\documentclass[aps,prl,twocolumn,superscriptaddress,nofootinbib,preprintnumbers]{revtex4}
\usepackage{color,graphicx,epsfig}
\usepackage{ifpdf}
\usepackage{amsmath}
\usepackage{bm}
\usepackage{color}
\usepackage[english]{babel}
\usepackage{graphicx}
\usepackage{amsfonts}
\usepackage{amssymb}
\usepackage{braket}
\usepackage{hyperref}
\usepackage{enumerate}
\usepackage{capt-of}
\usepackage[caption=false]{subfig}
\usepackage{tabularx}
\usepackage{xspace}

\usepackage{array, booktabs, makecell, multirow}

\definecolor{nicered}{rgb}{0.7,0.1,0.1}
\definecolor{nicegreen}{rgb}{0.1,0.5,0.1}
\hypersetup{colorlinks,citecolor=nicegreen,linkcolor=nicered,urlcolor=nicered}

\newcommand{\madgraph}{\texttt{MadGraph}\xspace}
\newcommand{\openloops}{\texttt{OpenLoops}\xspace}

\newcommand{\pysecdec}{\texttt{pySecDec}\xspace}

\newcommand{\disteval}{\texttt{disteval}\xspace}

\newcommand{\be}{\begin{equation}}
\newcommand{\ee}{\end{equation}}
\newcommand{\bea}{\begin{eqnarray}}
\newcommand{\eea}{\end{eqnarray}}
\definecolor{Red}{rgb}{1.,0.,0.}
\definecolor{randomcolor}{rgb}{0.2,0.5,0.7}

\DeclareMathAlphabet\mathbfcal{OMS}{cmsy}{b}{n}

\def\OMIT#1{}

\allowdisplaybreaks[1]

\begin{document}

\def\MSU{Department of Physics and Astronomy, Michigan State University, East Lansing, Michigan 48824, USA}
\def\UR{Institut für Theoretische Physik, Universit\"at Regensburg, 93040 Regensburg, Germany}
\def\DU{Institute for Particle Physics Phenomenology, Durham University, Durham DH1 3LE, UK}
\def\KIT{Institute for Theoretical Physics, Karlsruhe Institute of Technology (KIT), D-76131 Karlsruhe, Germany}

\preprint{IPPP/24/16, KA-TP-05-2024, MSUHEP-24-005, P3H-24-023}

\title{Complete Next-to-Leading Order QCD Corrections to $ZZ$ Production in Gluon Fusion}

\author{Bakul Agarwal}            
\email[Electronic address: ]{bakul.agarwal@kit.edu}
\affiliation{\KIT}
\affiliation{\MSU}

\author{Stephen Jones}            
\email[Electronic address: ]{stephen.jones@durham.ac.uk}
\affiliation{\DU}

\author{Matthias Kerner}            
\email[Electronic address: ]{matthias.kerner@kit.edu}
\affiliation{\KIT}

\author{Andreas von Manteuffel}            
\email[Electronic address: ]{manteuffel@ur.de}
\affiliation{\UR}
\affiliation{\MSU}

\begin{abstract}
We calculate the complete next-to-leading order (NLO) QCD corrections to loop-induced $gg \rightarrow ZZ$ production including full top-quark mass effects.
The two-loop virtual corrections are obtained by combining analytic results for the massless, Higgs-mediated, and one-loop factorizable contributions with numerically computed amplitudes containing the top-quark mass.
We show that the choice of subtraction scheme for the virtual contribution impacts the precision with which the virtual contribution must be evaluated in order to obtain sufficiently precise phenomenological predictions.
For direct production through a massive top-quark loop, we observe that the relative NLO corrections are large.
The direct massive and Higgs-mediated contributions individually increase relative to the massless production at high diboson invariant mass, but interfere destructively with each other.
At the Large Hadron Collider, the NLO corrections to the gluon channel give a sizable contribution to the $pp \to ZZ+X$ cross section at N${}^3$LO.
\end{abstract}

\maketitle
\section{Introduction}
\label{sec:introduction}
The advent of the High-Luminosity LHC brings exciting opportunities to measure Standard Model parameters at unprecedented precision.
An important process in this regard is $Z$ pair production, which is relevant for new physics searches~\cite{ATLAS:2020tlo,ATLAS:2020fry,CMS:2021itu,CMS:2019qem} and provides a significant background to Higgs production in the four-lepton channel~\cite{Aaboud:2018puo,Sirunyan:2018sgc,Sirunyan:2019twz,ATLAS:2020wny}, both for on- and off-shell Higgs bosons.
$Z$ boson pair production has been measured at $13.6$ TeV~\cite{ATLAS:2023dew} and used to constrain anomalous CP-odd neutral triple gauge couplings~\cite{ATLAS:2023zrv}.
Comparing resonant and nonresonant Higgs production allows for an indirect probe of the Higgs width~\cite{Caola:2013yja,Campbell:2013una}, and continuum $Z$ pair production can contribute significantly to off-shell Higgs production through interference~\cite{Kauer:2012hd,Kauer:2013qba}.
Constraints on the Higgs width have been obtained in this way both by CMS~\cite{CMS:2022ley} and ATLAS experiments~\cite{ATLAS:2023dnm}.
Given the phenomenological relevance, precise theoretical predictions for this process are desirable.

The gluon fusion channel for $Z$ pair production is loop-induced and starts to contribute to the hadronic process $pp \rightarrow ZZ+X$ only at next-to-next-to-leading order (NNLO).
Owing to high gluon luminosity at the LHC, this channel accounts for O(60\%) of the total NNLO correction~\cite{Cascioli:2014yka}.
NLO corrections to this channel are formally next-to-next-to-next-to-leading order with respect to the hadronic process.
In general, different partonic channels mix at higher orders in the perturbative expansion such that the concept of corrections to a specific subprocess is not well defined.
In the present case, however, one can define NLO corrections to the gluon channel by considering only the contributions in which neither of the $Z$ bosons couples to external quark lines.
The so-defined NLO corrections are indeed significant; estimates~\cite{Caola:2015psa,Grazzini:2018owa} give an overall O(5--8\%) increase to the total $pp \rightarrow ZZ$ cross section.

For the two-loop amplitudes, the massless corrections were computed a while back in~\cite{vonManteuffel:2015msa,Caola:2015ila}, while the top-quark contributions remained a challenge until recently, preventing the calculation of the exact NLO QCD corrections in this channel so far.
The top-quark contributions are expected to be important, particularly in the high invariant mass region for the production of longitudinal $Z$ bosons due to the Goldstone boson equivalence theorem~\cite{Lee:1977eg,Chanowitz:1985hj} and the interplay with Higgs-mediated production.
Approximate NLO corrections to $gg\to ZZ$ have been presented in~\cite{Melnikov:2015laa,Caola:2016trd,Campbell:2016ivq,Grober:2019kuf,Davies:2020lpf} and supplemented with a parton shower in~\cite{Alioli:2021wpn,Buonocore:2021fnj}.
Recently, the two-loop amplitudes have been calculated with full top-quark mass effects~\cite{Agarwal:2020dye,Bronnum-Hansen:2021olh}.

In this Letter, we take the next step and calculate the complete NLO QCD corrections to $gg\rightarrow ZZ$ with full top-quark mass effects.
Specifically, we consider all contributions to the cross section involving closed massless or massive quark loops to which the external $Z$ bosons couple either directly or through a Higgs boson.
For the real radiation contributions, we take into account also $gq$, $g\bar{q}$ and $q\bar{q}$ in addition to $gg$ initial states.
Using a fully differential setup, we compute the inclusive cross section and the diboson invariant mass distribution for $ZZ$ production at the LHC.

\label{sec:calculation}
\section{Details Of The Calculation}
The differential cross section for $gg\to ZZ$ at NLO can be written as
$
    \mathrm{d} \sigma_{\mathrm{NLO}} =  \mathrm{d}  \sigma_{\mathrm{B}} +  \mathrm{d}  \sigma_{\mathrm{V}} +  \mathrm{d} \sigma_{\mathrm{R}}\,,
$
where $\mathrm{d} \sigma_{\mathrm{B}}, \,\mathrm{d} \sigma_{\mathrm{V}}, \,\text{and} \,\,\mathrm{d} \sigma_{\mathrm{R}}$ correspond to Born, virtual, and real contributions, respectively.
In our calculation, we include effects due to $n_f=5$ massless quark flavors, a massive top quark, and a massive Higgs boson.
The calculation of the different contributions is described below.

\label{sec:virtuals}
\subsection{Born and virtual contributions}
We consider the partonic process
\begin{align}
\label{eq:ggzz}
g(p_1)\,+\,g(p_2)\:\xrightarrow{}\:Z(p_3)\,+\,Z(p_4)\, ,
\end{align}
for on-shell momenta, i.e.\ $p_1^2 = p_2^2 = 0$ and $p_3^2 = p_4^2 = m_Z^2$.

\begin{figure}[t]
    \setlength{\tabcolsep}{2ex}
    \begin{tabular}{cc}
    {\includegraphics[width=0.35\linewidth,height=2.5cm]{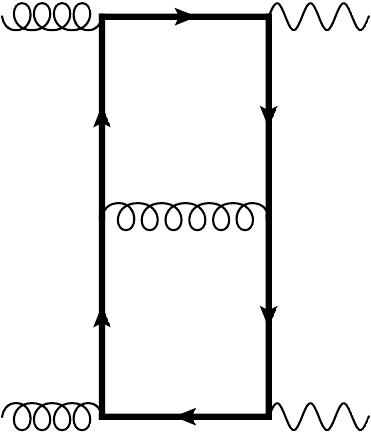}}
    &
    {\includegraphics[width=0.35\linewidth,height=2.5cm]{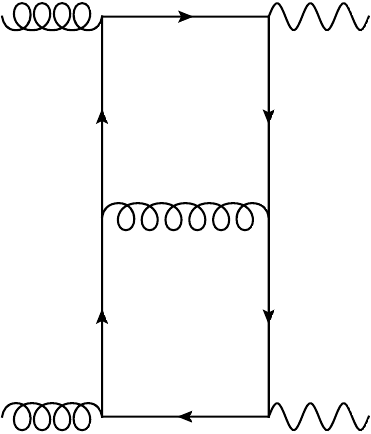}}
    \\
    A${}_h$  & A${}_l$
    \\[2ex]
    {\includegraphics[width=0.35\linewidth,height=2.5cm]{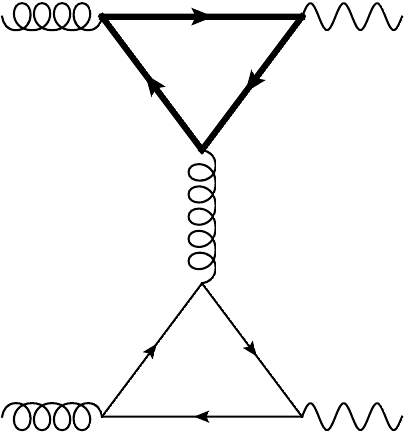}}
    &
    \raisebox{.35\height}{\includegraphics[width=0.35\linewidth,height=1.5cm]{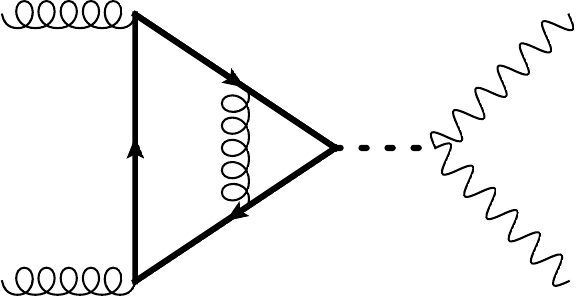}}
    \\
    B & C
    \end{tabular}
    \caption{Representative two-loop diagrams with closed quark loops entering the virtual corrections to $ZZ$ production. Thick lines denote top quarks.}
    \label{fig:diagramclasses}
\end{figure}

The one-loop amplitudes relevant for the Born contribution were calculated long ago in Refs.~\cite{Dicus:1987dj,Glover:1988rg,Zecher:1994kb}, while the two-loop corrections employed for this Letter were completed only recently.
We distinguish between different classes of contributions to the amplitude, depending on the couplings of the external $Z$ bosons.
Fig. \ref{fig:diagramclasses} shows a representative two-loop Feynman diagram for each of the following classes.

\textbf{Class A${}_h$:} Both $Z$ bosons couple directly to the same heavy top-quark loop.
For these one- and two-loop contributions, we use the recent calculation \cite{Agarwal:2020dye} by some of us that employed
a combination of syzygy techniques~\cite{Gluza:2010ws,Schabinger:2011dz,Ita:2015tya,Larsen:2015ped,Boehm:2017wjc,Agarwal:2020dye}, finite field methods~\cite{vonManteuffel:2014ixa,Peraro:2016wsq}, multivariate partial fractioning~\cite{Abreu:2019odu,Boehm:2020ijp,Chawdhry:2020for,Heller:2021qkz,Agarwal:2021grm}, and constructions of finite integrals, and the resulting finite master integrals were evaluated numerically with \pysecdec~\cite{Borowka:2018goh,Heinrich:2021dbf,Heinrich:2023til}.

\textbf{Class A${}_l$:} Both $Z$ bosons couple directly to the same light quark loop.
Analytical expressions for these one- and two-loop contributions were provided in~\cite{vonManteuffel:2015msa}, based on solutions for the master integrals~\cite{Gehrmann:2015ora} in terms of multiple polylogarithms.
We employ their implementation in the \texttt{VVamp} library and numerically evaluate the multiple polylogarithms using the code of~\cite{Vollinga:2004sn} included in \texttt{GiNaC}~\cite{Bauer:2000cp}.

\textbf{Class B:} The $Z$ bosons couple to different closed quark loops, each of which can be a light or a heavy quark.
At two loops, these corrections are one-particle reducible products of one-loop triangles.
These are the only diagrams involving Dirac traces with an odd number of $\gamma_5$ matrices and contributions related to the chiral anomaly can arise due to a mass splitting within a weak isodoublet.
Consequently, one should consider them for sums over a complete quark generation, and for our calculation with five massless quarks just the third generation contributes due to $m_b \neq m_t$.
These contributions have been presented in~\cite{Campbell:2016ivq}; we use the recalculation in~\cite{Agarwal:2020dye} for this work.

\textbf{Class C:} The $Z$-boson pair is produced via the decay of an intermediate off-shell Higgs boson, which couples to a heavy quark loop.
We employ an in-house implementation of these Higgs-mediated contributions based on the differential equations approach, similar to the calculation in~\cite{Anastasiou:2006hc}.
In the high invariant mass region, above the top-quark threshold, one finds interesting interferences between the Higgs-mediated and direct production of longitudinally polarized $Z$ bosons.
At one loop, it has been discussed in~\cite{Glover:1988rg} that the interference is destructive and exhibits a cancellation of the leading term at high energy, as required by the unitarity of the $t\bar{t}\to ZZ$ subprocess;
we observe a strong destructive interference also at two loops.

After UV renormalization and IR subtraction, details of which are provided in Refs.~\cite{vonManteuffel:2015msa,Agarwal:2020dye},
the finite remainders for the helicity amplitudes can be written as
\begin{align}
    \mathcal{M}_{\lambda}^{\mathrm{fin}} = \left(\frac{\alpha_s}{2\pi}\right) \mathcal{M}_{\lambda}^{(1)} + \left(\frac{\alpha_s}{2\pi}\right)^2 \mathcal{M}_{\lambda}^{(2)} + \mathcal{O} \left( \alpha_s^3 \right),
\end{align}
where $\lambda = \{\lambda_1, \lambda_2, \lambda_3, \lambda_4\}$ specify the polarizations of the external particles. 
Here,
$\mathcal{M}^{(1)}_
\lambda,\mathcal{M}^{(2)}_\lambda$ are the one- and two-loop finite remainders constructed from the form factor decomposition described in~\cite{vonManteuffel:2015msa}.

We define the squared one-loop amplitude $\mathcal{V}^{(1)}$ as
\begin{align}
 \label{eq:vfin}
     \mathcal{V}^{(1)} &=  \frac{1}{N}\, \sum_{\lambda,\mathrm{color}} \mathcal{M}_{\lambda}^{*(1)} \, \mathcal{M}_{\lambda}^{(1)} \,,
\end{align}
where we divide by $N=2^2 \times 8^2\times 2$ to account for the averaging over spins and colors in the initial state and the symmetry factor due to identical particles in the final state.
To optimize the sampling of the two-loop amplitude for our full result, we separate it according to the classes defined above
\begin{align}
\mathcal{M}^{(2)}_\lambda = \mathcal{M}^{(2)}_{\lambda,A_h} + \mathcal{M}^{(2)}_{\lambda,A_l} + \mathcal{M}^{(2)}_{\lambda,B} + \mathcal{M}^{(2)}_{\lambda,C}\,\,.
\end{align}

The interference with the full Born amplitude $\mathcal{M}^{(1)}_\lambda$ can then be written as
\begin{align}
\mathcal{V}^{(2)} &=\frac{1}{N}\, \sum_{\lambda,\mathrm{color}} \,2\, \mathrm{Re} \left( \mathcal{M}^{*(1)}_{\lambda} \, \mathcal{M}^{(2)}_{\lambda,A_h} \right. \nonumber \\
& \left. + \mathcal{M}^{*(1)}_{\lambda} \, (\mathcal{M}^{(2)}_{\lambda,A_l} + \mathcal{M}^{(2)}_{\lambda,B} + \mathcal{M}^{(2)}_{\lambda,C})  \right).
\end{align}
We sample the first interference term using massive virtual amplitude evaluations distributed according to unweighted events based on the top-quark only Born distribution, as described below.
The remaining Born-virtual interference terms are evaluated with higher statistics: they are sampled dynamically by our phase-space generator code using the VEGAS algorithm~\cite{Lepage:1977sw,Hahn:2004fe}, with an additional weight to ensure that sufficiently many events are generated at high invariant mass.

By default, we use the $q_T$ subtraction scheme~\cite{Catani:2013tia} for our two-loop amplitudes and the Catani-Seymour scheme~\cite{Catani:1996vz} to construct the dipoles for our real radiation diagrams.
This mismatch amounts to a term proportional to the Born contribution which is straightforward to add (subtract) from the virtual (real) contributions as we desire.
Similarly, we can obtain the virtual and real contributions in the original Catani subtraction scheme~\cite{Catani:1998bh} by adding terms proportional to the Born contribution.
Starting with the I operator in the $q_T$ scheme, we can obtain results in the Catani and Catani-Seymour subtraction schemes by adding $2\mathrm{Re}(\Delta\mathbf{I})\cdot \mathcal{V}^{(1)}$ to $\mathcal{V}^{(2)}$, where
\begin{align}
    \Delta\mathbf{I}_{\mathrm{C}} &= -\frac{1}{2} \pi^2 C_A + i \pi \beta_0, \\
    \Delta\mathbf{I}_{\mathrm{CS}} &= -i\pi C_A \left[ \frac{1}{\epsilon} + \ln \left( \frac{\mu_R^2}{\hat{s}} \right) \right] - \frac{\pi^2 C_A}{3} + \beta_0 + k_g,
\end{align}
with the Mandelstam invariant $\hat{s}=(p_1+p_2)^2$, $\beta_0 = (11/6) C_A - (2/3) T_F n_f$ and $k_g = (67/18 - \pi^2/6)\, C_A - (10/9) T_F n_f$.
The imaginary parts of the shifts do not contribute when computing the Born-virtual interference. 
We utilize these conversions to construct our virtual and real contributions in all three schemes ($q_T$, Catani-Seymour, and Catani).

The calculation of the class A${}_{h}$ massive virtual amplitudes was described in Ref.~\cite{Agarwal:2020dye}.
In the current work, the integrated virtual contribution is obtained using a new implementation of these amplitudes, relying on the distributed evaluation (\disteval) feature of \pysecdec, developed in tandem with this work.
We evaluate 3000 unweighted events (distributed according to the Born cross section) on multiple GPUs.
A major improvement, required to obtain sufficiently stable results in all phase-space regions, is the evaluation of the two-loop amplitudes directly in the physical helicity basis (not the $A$-basis form factors presented in Refs.~\cite{Agarwal:2020dye,vonManteuffel:2015msa}), for which we set a relative accuracy target of 1$\%$ per amplitude.
On a test grid of phase-space points, taken from Ref~\cite{Agarwal:2020dye}, the mean evaluation time per phase-space point is 1.9 min with the new implementation and accuracy goal compared to 4.8 h with the original code.
Our new virtual amplitude code is checked against our original code and an independent calculation using different methods described in Ref.~\cite{Bronnum-Hansen:2021olh}.

\label{sec:reals}
\subsection{Real contributions}
\begin{figure}[t]
    \setlength{\tabcolsep}{2ex}
    \begin{tabular}{cc}
    {\includegraphics[width=0.35\linewidth,height=1.5cm]{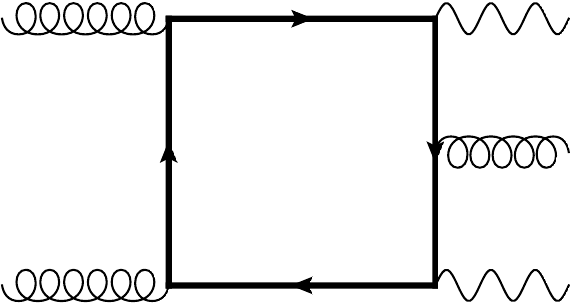}}
    &
    {\includegraphics[width=0.35\linewidth,height=1.8cm]{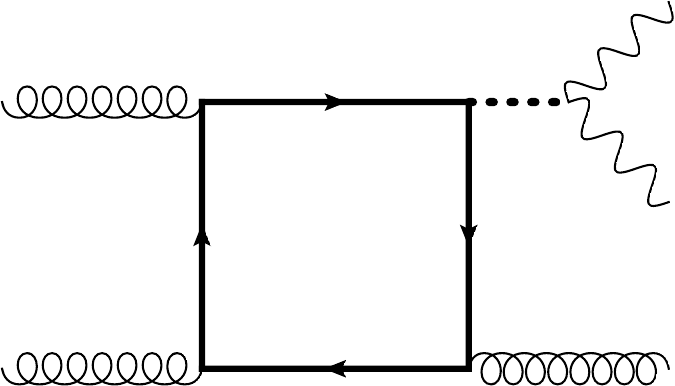}}
    \\[2ex]
    {\includegraphics[width=0.35\linewidth,height=2cm]{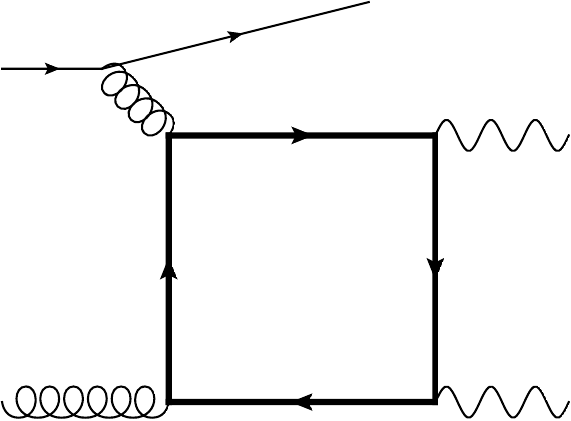}}
    &
    {\includegraphics[width=0.35\linewidth,height=2cm]{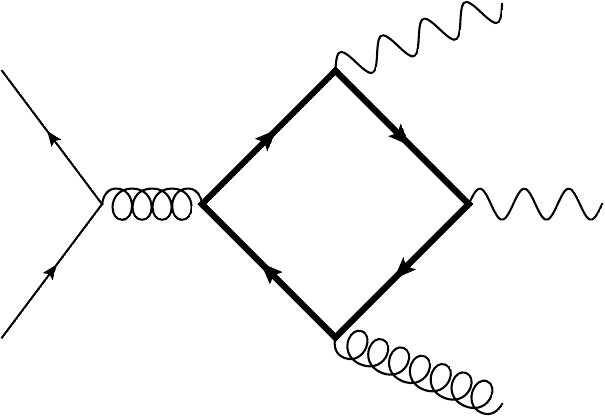}}
    \end{tabular}
    \caption{Representative diagrams with a closed quark loop entering the real corrections to $ZZ$ production. Different partonic channels and loops with light or top quarks contribute.}
    \label{fig:realdiagrams}
\end{figure}

The real corrections for this process include the partonic channels
$gg\rightarrow ZZ+g$, $q(\bar{q})\,g\rightarrow ZZ+q(\bar{q})$, and $q\bar{q}\rightarrow ZZ+g$;
representative Feynman diagrams are shown in Fig.~\ref{fig:realdiagrams}.
For the real radiation diagrams, we require that both $Z$ bosons are coupled to the closed fermion loop; in particular, we exclude diagrams that involve a $Z$ boson coupling to an incoming or outgoing quark line.

The amplitudes for the real radiation diagrams are generated using \texttt{GoSam}~\cite{Cullen:2011ac,GoSam:2014iqq} and validated numerically at the level of individual phase-space points against \madgraph~\cite{Alwall:2014hca,Frederix:2018nkq} and \openloops~\cite{Buccioni:2019sur} (which uses \texttt{COLLIER}~\cite{Denner:2016kdg}, \texttt{CutTools}~\cite{Ossola:2007ax}, and \texttt{OneLOop}~\cite{vanHameren:2010cp}). 
Our integrated real matrix elements are also validated against \madgraph by computing the LO $ZZ$ plus jet cross section with several different values of $p_{T,j}$ cut.
To obtain stable numerical results we use the Ninja integrand reduction package~\cite{Mastrolia:2012bu,Peraro:2014cba} with a quadruple precision rescue system. 
Our rescue system employs a three-step procedure: each point is evaluated twice in double precision after an azimuthal rotation about the beam axis; if the double results do not agree to 11 digits then the amplitude is reevaluated in quadruple precision, if the quadruple evaluation does not agree with the double results within eight digits then a second rotated quadruple result is obtained. 
The result is discarded if the two quadruple results do not agree within 11 digits,
affecting $\sim 10^{-5}$ of the points. 
The most common feature of the rejected points is the presence of a soft jet.
Soft and collinear singularities arising from the phase-space integration are cancelled locally by Catani-Seymour dipoles.

To obtain the Catani-Seymour dipole contributions, we compute the required spin-correlated Born matrix elements using the massless quark amplitudes from the \texttt{VVamp} library~\cite{vonManteuffel:2015msa} and an in-house implementation of the one-loop massive quark amplitudes using \texttt{LoopTools}~\cite{vanOldenborgh:1989wn,Hahn:1998yk} to evaluate the one-loop master integrals.
For the calculation of the dipoles, we observed that the spin-correlated Born matrix elements were numerically unstable in highly soft or collinear regions using the form factors of~\cite{Agarwal:2020dye}. 
Switching to the orthogonal form factors defined in~\cite{Davies:2020lpf} improved numerical stability, and much better cancellation in the singular regions was observed.

\section{Results}
In this section, we present our NLO results for $ZZ$ production in the gluon channel for the LHC.
We set $G_F = 1.1663787 \cdot 10^{-5} \ \mathrm{GeV}^{-2},\, m_Z = 91.1874\ \mathrm{GeV},\, m_W = 80.2959\ \mathrm{GeV},$ and $m_t = 173.016\ \mathrm{GeV}.$
For the calculation of the two-loop massive amplitudes, we fix the ratio $m_Z^2/m_t^2=5/18$ as described in~\cite{Agarwal:2020dye}.
We set $\mu_R=\mu_f=m_{ZZ}/2$ as central values for renormalization and factorization scales respectively, and use the $\mathrm{CT18NLO}$~\cite{Hou:2019efy} parton distribution functions interfaced via \texttt{LHAPDF}~\cite{Buckley:2014ana} to calculate the total cross section.
Uncertainty estimates are obtained by simultaneously varying the renormalization and factorization scales by a factor of 2 around the central scale $\mu = m_{ZZ}/2$.

\begin{figure}
    \centering
    \includegraphics[width=1.05\linewidth]{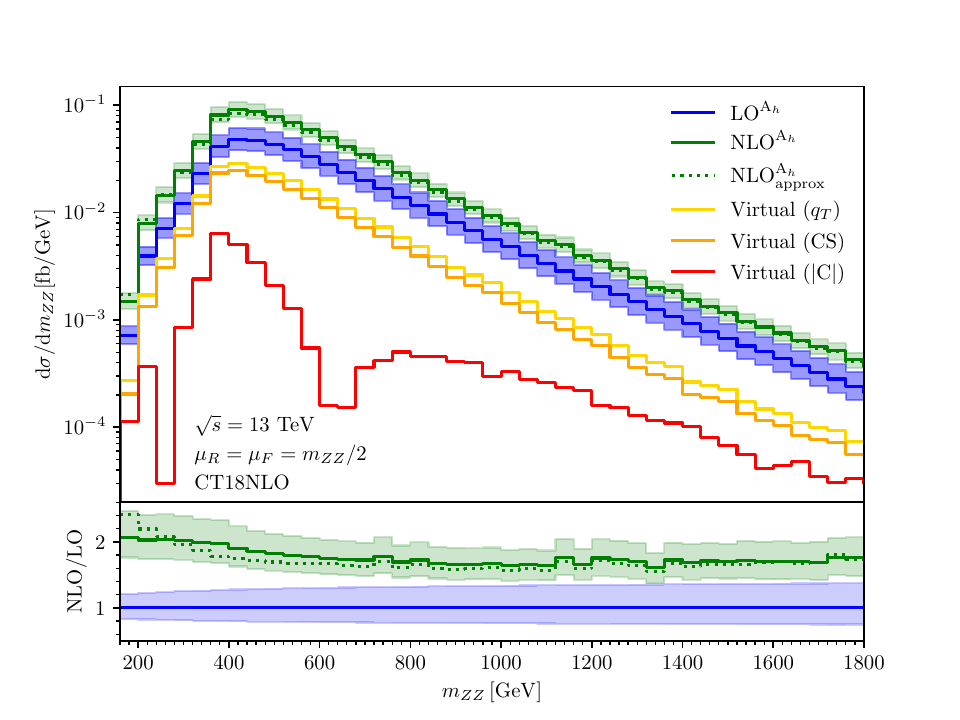}
    \caption{Top-quark-only contributions to the $ZZ$ invariant mass distribution in $pp$ collisions.
    The absolute value of the two-loop virtual correction is shown separately in the $q_T$, Catani-Seymour (CS), and Catani (C) schemes.
    The dashed curve represents an approximate NLO result obtained by rescaling the massive Born amplitude with the massless K-factor.}
    \label{fig:mzztoponly}
\end{figure}

In Fig.~\ref{fig:mzztoponly}, we show the invariant mass distribution for this process considering only diagrams of class A${}_h$, i.e.\ 
contributions involving both $Z$ bosons coupled to the same closed top-quark loop.
The shaded bands indicate the scale uncertainty.
We find that the massive NLO corrections are large, enhancing the top-only cross section by a factor of $1.8$ at $\sqrt{s}=13$ TeV hadronic center-of-mass energy.
The NLO corrections to the $Z$-boson invariant mass distribution are large but rather flat; they amount to a factor 2 increase in the distribution near the $ZZ$ production threshold and decrease to a factor of around 1.7 at $1$ TeV.
For the corresponding integrated cross section at $\sqrt{s} = 13\, \mathrm{TeV}$ we obtain,
$\sigma_{\mathrm{LO}}^{\mathrm{A}_h} = 19.00^{+29.4\%}_{-21.4\%}$ fb,
$\sigma_{\mathrm{NLO}}^{\mathrm{A}_h} = 34.46(6)^{+16.4\%}_{-14.4\%}$ fb, where the number in parenthesis indicates the statistical Monte Carlo error.

In Fig.~\ref{fig:mzztoponly}, we also show the subtracted virtual contribution obtained using different IR subtraction schemes.
The NLO result does not depend on the choice of subtraction scheme, with different schemes amounting to reshuffling contributions proportional to $\mathcal{V}^{(1)}$ between the subtracted reals and virtuals. 
However, we do observe that while the virtuals in the $q_T$ and Catani-Seymour schemes behave rather similarly, they are heavily suppressed in the Catani scheme.
The shape of the Catani scheme virtual corrections is similar to the other schemes, but, the corrections are shifted down and are negative in the bins $m_{ZZ} < 220 \ \mathrm{GeV}$ and $m_{ZZ} > 620 \ \mathrm{GeV}$.
This implies that in the Catani scheme, the majority of the NLO correction comes from the IR subtracted real contribution and the subtracted virtual contribution accounts for less than $2\%$ of the total cross section.
We point out that the finite remainders in the Catani scheme were previously shown to be more sensitive to kinematic expansions of the two-loop expressions than in the $q_T$ scheme~\cite{Agarwal:2020dye}, and may thus be interpreted as representing more directly the genuine two-loop effects.
Choosing a scheme where the virtuals are numerically small can be of practical importance in situations where their exact evaluation is possible but computationally expensive, since one can reduce the number of phase-space points for the numerical integration in this way.
Nevertheless, in the present work, we were able to obtain sufficient statistics that the virtuals could be reliably obtained in each subtraction scheme, as shown.

We also compare our results to an approximation, NLO$_\mathrm{approx}^{\mathrm{A}_h}$ similar to~\cite{Grazzini:2018owa}, obtained using exact ingredients except for the massive two-loop virtual amplitudes, which are replaced by the top-quark only Born amplitude rescaled by the ratio $\frac{1}{2} \mathcal{V}^{(2)}/\mathcal{V}^{(1)}$ computed using only the massless quark amplitudes.
This rescaling is performed fully differentially at the level of individual phase-space points.
We find that the approximation describes the exact results well in most of the phase-space for the unpolarized distributions, particularly in the high energy region.

In other loop-induced processes, an uncertainty related to the choice of the top-quark mass renormalization scheme and scale is known to be large at NLO, dominating the QCD scale uncertainty~\cite{Baglio:2018lrj,Baglio:2020wgt,Chen:2022rua,Degrassi:2022mro}.
To estimate this uncertainty, we can compare the result in the on-shell (OS) scheme with a result in the $\overline{\mathrm{MS}}$ scheme at the scale $\mu_t = 2 m_t^\mathrm{OS}$.
Converting the top-quark mass to the $\overline{\mathrm{MS}}$ scheme we obtain $m_t(\mu_t) = 154.6\ \mathrm{GeV}$~\cite{Chetyrkin:2000yt,Herren:2017osy}.
Considering $\sigma_{\mathrm{LO}}^{\mathrm{A}_h}$ only, at LO we find that the $\overline{\mathrm{MS}}$ result is 8.7\% larger than the OS result.

In our calculation, the ratio of the mass of the $Z$ boson and the top quark is fixed to the OS value in the two-loop virtual correction, preventing the direct evaluation of the scheme uncertainty at NLO.
To estimate the size of the uncertainty, we can renormalize the top-quark mass in the $\overline{\mathrm{MS}}$ scheme and vary the mass in all parts of the calculation except the finite $\mathcal{O}(\epsilon^0)$ term of the bare two-loop amplitude.
We find that at NLO the $\overline{\mathrm{MS}}$ result is $4.1 \pm 0.4 \%$ bigger than the OS result. 
The stability of our estimate is assessed by computing it in both the Catani and $q_T$ subtraction schemes.

\begin{figure}
    \centering
    \includegraphics[width=1.05\linewidth]{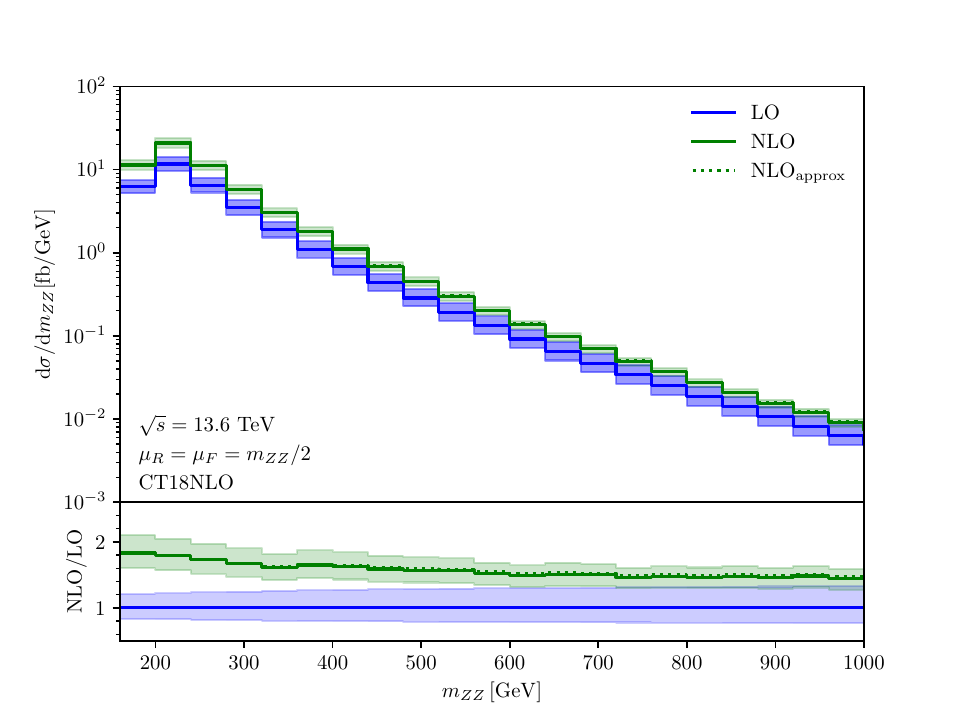}
    \caption{
    Diboson invariant mass distribution for gluon-initiated $ZZ$ production at the LHC.
    The solid curves represent the LO and NLO results with complete massless and massive contributions, including Higgs-mediated diagrams.
    The dashed curve represents an approximate NLO result obtained as described in the text.}
    \label{fig:mzzfull}
\end{figure}

In Fig.~\ref{fig:mzzfull}, we show the invariant mass distribution for $ZZ$ production in the gluon channel for the LHC with $\sqrt{s}=13.6$~TeV, taking into account all massless and massive contributions, including those mediated by a Higgs boson.
As above, the shaded bands indicate the scale uncertainty.
We find that the complete NLO corrections are large, ranging from 1.8 near the $ZZ$ production threshold and dropping to around 1.4 at 1~TeV invariant mass.

For the dashed curve, NLO${}_\mathrm{approx}$, we again employ the approximation in which the two-loop massive virtual amplitude is replaced by the rescaled top-quark only Born amplitude, as described above.
At low invariant mass, the cross section is dominated by diagrams containing loops of massless quarks and, to a lesser extent, their interference with the Higgs-mediated contribution, both of which are included exactly in the approximation.
Conversely, at high invariant mass, where the massive contribution is more important, the massive amplitudes are approximated well.
As a result, we observe that the approximation works well across the entire invariant mass range for the full unpolarized NLO correction.

For the full NLO cross section in the gluon channel at $\sqrt{s}=13.6\,\mathrm{TeV}$ with exact dependence on the top-quark mass, we obtain
\begin{align}
    \sigma_{\mathrm{LO}} &= 1316^{+23.0\%}_{-18.0\%}\,\mathrm{fb}\,,\\
    \sigma_{\mathrm{NLO}} &= 2275(12)^{+14.0\%}_{-12.0\%}\,\mathrm{fb}\,.
\end{align}
%
Here, the number in parenthesis indicates the statistical Monte Carlo error, while the percentages specify the uncertainty stemming from simultaneous variation of the renormalization and factorization scales by a factor of 2.
The complete NLO corrections to the gluon channel are large, increasing the contribution by a factor of 1.7 compared to the leading order and beyond the naive scale uncertainty estimate.
The corrections approximately half the scale uncertainty.
The impact of including the direct massive and Higgs-mediated contributions at the level of the total cross section is moderate, decreasing the NLO cross section by around 2$\%$ compared to purely massless contributions.

\section{Conclusions}
We have presented a complete calculation of the NLO QCD corrections to $gg\rightarrow ZZ$ retaining full top-quark mass effects.
For the LHC at center-of-mass energy $\sqrt{s}=13.6$~TeV, we find that the total cross section of the gluon-induced channel is enhanced by a factor of 1.7 relative to the LO correction.
The diboson invariant mass distribution $K$ factor falls from around 1.8 at production threshold to 1.4 at 1~TeV.

We show that the choice of infrared subtraction scheme has a great impact on the accuracy with which the two-loop finite remainders need to be known.
In particular, we find that in Catani's original subtraction scheme, the virtual contributions are numerically suppressed.
This observation can help to reduce the computational cost for sampling these challenging contributions and should be taken into account when quantifying the quality of approximations for the finite remainders.
Our analysis also shows that an approximation of the two-loop massive amplitudes similar to~\cite{Grazzini:2018owa} works well at the level of the full unpolarized NLO invariant mass distribution.

In future work, it would be interesting to study the impact of top-quark mass effects on other differential observables and in the presence of anomalous couplings, e.g., $Zt\bar{t}$~\cite{Azatov:2016xik,Cao:2020npb}.
We observe that at high invariant mass there is strong destructive interference at LO as well as NLO between diagrams in which the $Z$ bosons couple directly to a massive quark line and those in which the $Z$ bosons are produced via an intermediate Higgs boson; this cancellation could be spoiled by such anomalous couplings.
The corrections presented here contribute at N${}^3$LO to the hadronic cross section, and should be combined with the quark-initiated contributions at least to NNLO for such phenomenological applications.
\\


\begin{acknowledgments}
\textbf{Acknowledgments}
We would like to thank Gudrun Heinrich, Stephan Jahn, and C.-P.~Yuan for useful discussions and related work.
This research was supported by the Deutsche Forschungsgemeinschaft (DFG, German Research Foundation) through grant 396021762 - TRR 257, the National Science Foundation through grant 2013859, the Royal Society through the University Research Fellowship grant URF/R1/201268, and the UK Science and Technology Facilities Council through grant ST/X000745/1.
We gratefully acknowledge support and resources
provided by the the High Performance Computing Center (HPCC) at Michigan State University.
Our Feynman diagrams were generated using {\tt JaxoDraw} \cite{Binosi:2003yf}, based on {\tt AxoDraw} \cite{Vermaseren:1994je}.

\end{acknowledgments}

\bibliographystyle{bibliostyle}
\bibliography{biblio}

\newpage
\onecolumngrid

\makeatletter
\renewcommand\@biblabel[1]{[#1S]}
\makeatother

\end{document}